\documentclass[aps,prd,preprint,floatfix]{revtex4}
\usepackage{epsfig}
\usepackage{bm}
\usepackage{dcolumn}

\begin{document}
   
\preprint{\rightline{ANL-HEP-PR-06-48}}
   
\title{The RHMC algorithm for theories with unknown spectral bounds.}
   
\author{J.~B.~Kogut}\thanks{Supported in part by NSF grant NSF PHY03-04252.}
\affiliation{Department of Energy, Division of High Energy Physics, Washington,
DC 20585, USA}
 \author{\vspace{-0.2in}{\it and}}
\affiliation{Dept. of Physics -- TQHN, Univ. of Maryland, 82 Regents Dr.,
College Park, MD 20742, USA}
\author{D.~K.~Sinclair}\thanks{This work was supported by the U.S.
Department of Energy, Division of High Energy Physics, Contract \\*[-0.1in]
W-31-109-ENG-38.}
\affiliation{HEP Division, Argonne National Laboratory, 9700 South Cass Avenue,
Argonne, IL 60439, USA}
                                                                                
\begin{abstract}
The Rational Hybrid Monte Carlo (RHMC) algorithm extends the Hybrid Monte Carlo
algorithm for lattice QCD simulations to situations involving fractional
powers of the determinant of the quadratic Dirac operator. This avoids the
updating increment ($dt$) dependence of observables which plagues the
Hybrid Molecular-dynamics (HMD) method. The RHMC algorithm uses rational
approximations to fractional powers of the quadratic Dirac operator. Such
approximations are only available when positive upper and lower bounds to the
operator's spectrum are known. We apply the RHMC algorithm to simulations of 2
theories for which a positive lower spectral bound is unknown: lattice QCD
with staggered quarks at finite isospin chemical potential and lattice QCD
with massless staggered quarks and chiral 4-fermion interactions ($\chi$QCD).
A choice of lower bound is made in each case, and the properties of the RHMC
simulations these define are studied. Justification of our choices of lower
bounds is made by comparing measurements with those from HMD simulations, and
by comparing different choices of lower bounds.
\end{abstract}
                                                                                
\maketitle

\section{Introduction}

For lattice field theories with fermions, and in particular for lattice QCD,
where there exists a local action describing the dynamics with the desired
number of fermion flavours, the preferred simulation method is the hybrid
Monte-Carlo (HMC) algorithm \cite{Duane:1987de}. The HMC augments a hybrid
molecular-dynamics update with a discrete `time' step ($dt$), with a global
accept/reject step at the end of each trajectory. This accept/reject step
makes the algorithm exact, removing all $dt$ dependence from measured
observables.

For staggered and improved staggered fermions with numbers of flavours which
are not multiples of 4 and for other fermion schemes with odd numbers of
flavours, there is no local action, and one needs to take fractional powers of
the fermion determinant. Until recently, simulations of such theories used
hybrid molecular-dynamics simulations with `noisy' fermions, usually the R
algorithm (HMD(R)) \cite{Gottlieb:1987mq}. The problem is that in HMD(R)
simulations, the observables depend on $dt$, the leading departure from the
desired $dt=0$ results being ${\cal O}(dt^2)$. Hence one should either
simulate at several (small) $dt$ values and extrapolate to $dt=0$, or perform
simulations at such a small $dt$ value that the ${\cal O}(dt^2)$ errors are
negligible.

In the RHMC algorithm
\cite{Kennedy:1998cu,Clark:2003na,Clark:2004cp,Clark:2005sq,Clark}, instead of
taking fractional powers of the determinant of the quadratic Dirac operator,
one takes fractional powers of the operator itself. This defines an action
(nonlocal) and thus permits a global accept/reject step. Since the quadratic
Dirac operator is positive definite, such fractional powers are well defined
in terms of the eigenmodes of the operator. To make this algorithm practical,
one needs approximations to the fractional powers of the eigenvalues, which
can be applied directly to matrices. The RHMC algorithm uses rational
approximations and their partial-fraction expansions to approximate such
fractional powers over the range of eigenvalues of the quadratic Dirac
operator. Provided that the condition number (ratio of maximum to minimum
eigenvalues) of this operator is not too large, sufficiently good
approximations to this fractional power can be obtained with a modest order of
numerator and denominator polynomials in this approximation.

The first theory we consider is lattice QCD at finite isospin chemical
potential $\mu_I$, but with no explicit symmetry breaking parameter
\cite{Kogut:2004zg,Kogut:2005yu,sinclair}.
The spectrum of the quadratic Dirac operator for this theory has an unknown
lower bound. However, we can make a reasonable guess as to what is a
conservative estimate of the lower bound for the small $\mu_I$ values of
interest. The need for such an exact algorithm is indicated, since, in HMD(R)
simulations, the Binder cumulants which are used to determine the nature of
the finite temperature phase transition for this theory depend strongly on
$dt$ \cite{Kogut:2005yu,sinclair,Philipsen:2005mj,
deForcrand:2006pv}. We have adapted the RHMC algorithm to this theory with
3-fermion flavours and tested the reversibility needed for its implementation.
To check the validity of our speculative lower bound to the Dirac spectrum, we
have compared our measurements with extrapolations to zero $dt$ of our results
from earlier HMD(R) simulations at various $dt$s. In addition, we have performed
simulations using rational approximations which assume much smaller lower
bounds to the spectrum, at the largest $\mu_I$ value and selected $\beta$ and
$m$ values which we used with the original choice of bounds. Indications are
that our original choice of lower bounds are adequate.

The second theory we are considering is 2-flavour lattice QCD at zero quark
mass, using the $\chi$QCD action, which incorporates an irrelevant chiral
4-fermion interaction that allows simulations at zero quark mass
\cite{Kogut:1998rg,Kogut:2006gt}. Here again, we
have reasonable estimates of the upper bound of the spectrum of the quadratic
Dirac operator, but the lower bound is unknown. Again we make an educated
guess as to what is a reasonably conservative lower bound for the spectrum of
this operator and test our choice. For a $24^3 \times 8$ lattice, we found
that using 32-bit floating point precision, was the main limit on
reversibility. With 64-bit precision, the convergence criteria for the
multimass Dirac operator inversion was the limiting factor, leading us to
prefer using 64-bit precision. We tried 3 different choices of speculative
lower bounds on the quadratic Dirac operator comparing the inversions at the
beginning and at the end of each trajectory. For the 2 lowest choices, these
inverses agreed to within the known accuracy of the rational approximations
used, for all trajectories in the run. The highest choice agreed with the other
2 for almost all trajectories in the run. In addition we made direct comparison
of the results of simulations using the RHMC algorithm with earlier
measurements using the HMD(R) algorithm on small ($8^3 \times 4$) lattices.

Section~2 introduces the 2 theories under consideration with rational 
approximations. In section~3 we discuss the case of QCD at finite $\mu_I$,
while in section~4 we consider $\chi$QCD. Finally in section~5 we present
discussions and conclusions.

\section{QCD at finite $\mu_I$, $\chi$QCD and rational approximations}

In this section we present two Lattice QCD actions for which the lower bound
on the spectrum of the quadratic Dirac operator is unknown. This is important
for the application of the RHMC algorithm for numbers of fermion flavours other
than multiples of 8. One can only find a rational approximation to a fractional
power $\alpha$ of a positive (semi-)definite matrix to any desired relative
accuracy, if one knows positive lower and upper bounds to its spectrum.

If $M$ is a positive (semi-)definite matrix with eigenvectors $|\lambda\rangle$
and corresponding (non-negative) eigenvalues $\lambda$, and $\alpha$ is a real 
number,
\begin{equation}
M^\alpha = \sum_\lambda \lambda^\alpha |\lambda\rangle\langle\lambda|
\end{equation}
gives a unique definition of any real power of $M$.
If one knows the spectral bounds on $M$ i.e. that $0 < a \le \lambda \le b$,
for some $a$ and $b$, then one can find a rational approximation to 
$\lambda^\alpha$ of the form
\begin{equation}
\lambda^\alpha \approx P(\lambda)/Q(\lambda)
\end{equation}
where $P$ and $Q$ are polynomials in $\lambda$, such that for a any given 
(small) positive $\epsilon$
\begin{equation}                                                               
\stackrel{\rm\textstyle Max}{\lambda \in [a,b]}
{|\lambda^\alpha - P(\lambda)/Q(\lambda)| \over |\lambda^\alpha|}
\le \epsilon.
\end{equation}
As $\epsilon \rightarrow 0$ or $a/b \rightarrow 0$, the orders of $P$ and/or
$Q$ needed $\rightarrow \infty$. The method of calculating the best
approximation for fixed orders of $P$ and $Q$ was specified by Remez. We use
an implementation provided by Clark and Kennedy \cite{clark2} (Note that this
approximation also yields the best rational approximation to
$\lambda^{-\alpha}$). This rational approximation to the eigenvalues of
$M^\alpha$ provides a rational approximation to $M^\alpha$ itself as
\begin{equation}
M^\alpha \approx P(M)Q^{-1}(M),
\label{eqn:Malpha}
\end{equation}
which can be implemented directly in terms of the matrix $M$ without having to
find its eigenvalues and eigenvectors. Expanding the right-hand-side of
equation~\ref{eqn:Malpha} in terms of partial fractions allows the action of
$M^\alpha$ on any vector to be calculated using a multishift conjugate
gradient algorithm \cite{Frommer:1995ik,Jegerlehner:1996pm} for little more
than the cost of a single conjugate gradient inversion.

\subsection{QCD at finite $\mu_I$}

Lattice QCD at finite isospin chemical potential $\mu_I$, has the staggered 
quark action \cite{Kogut:2004zg}
\begin{equation}
S_f=\sum_{sites} \left[\bar{\chi}[D\!\!\!\!/(\frac{1}{2}\tau_3\mu_I)+m]\chi
                   + i\lambda\epsilon\bar{\chi}\tau_2\chi\right],
\end{equation}
where $D\!\!\!\!/(\frac{1}{2}\tau_3\mu_I)$ is the standard staggered quark
transcription of $D\!\!\!\!/$ with the links in the $+t$ direction multiplied
by $\exp(\frac{1}{2}\tau_3\mu_I)$ and those in the $-t$ direction multiplied by
$\exp(-\frac{1}{2}\tau_3\mu_I)$. For finite temperature simulations with
$\mu_I < m_\pi$, we set the coefficient $\lambda$ of the explicit symmetry 
breaking term to zero. As it stands, this action describes 4 quark flavours
(in the continuum limit).

To tune this to $N_f$ flavours following the RHMC approach, we replace this
with the pseudo-fermion action
\begin{equation}
S_{pf}=p_\psi^\dag {\cal M}^{-N_f/8} p_\psi
\end{equation}
where $p_\psi$ is the momentum conjugate to the pseudo-fermion field $\psi$.
\begin{equation}
{\cal M} = [D\!\!\!\!/(\frac{1}{2}\tau_3\mu_I)+m]^\dag
           [D\!\!\!\!/(\frac{1}{2}\tau_3\mu_I)+m] + \lambda^2
\end{equation}
is the quadratic Dirac operator, and we set $\lambda=0$. To apply the RHMC
algorithm we will need rational approximations to ${\cal M}^{-N_f/8}$ and
${\cal M}^{\pm N_f/16}$.

It is easy to see that $[\cosh(\frac{1}{2}\mu_I)+3+m]^2$ is an upper bound to
the spectrum of ${\cal M}$. We, in fact, choose an upper bound of $25$ which
exceeds this bound for any choice of $\mu_I$ and $m$ we are ever likely to
consider. At $\mu_I=0$, $m^2$ is a lower bound on the spectrum. One might
expect that this bound is overly conservative, and that the actual lower bound
would be closer to $\frac{1}{2}m_\pi$. For $\mu_I > 0$, we do not know the
lower bound of the spectrum of the quadratic Dirac operator. However, the
effective pion mass is $m_\pi - \mu_I$, so if this gives at least some guide
as to the lower bound of this operator, we would expect the lower bound of its
spectrum to decrease smoothly as $\mu_I$ is increased, becoming zero at
$\mu_I=m_\pi$, the value of $\mu_I$ at the zero temperature transition to the
superfluid phase. Our observations from HMD(R) simulations, indicate that in
this low $\mu_I$ regime the Dirac operator becomes more singular as $\mu_I$ is
increased, in a controlled way. Since we are interested in quark masses $m \ge
0.02$, we choose a speculative lower bound of $1.0 \times 10^{-4}$, 4 times
lower than the known lower bound at $\mu_I=0$ for $m=0.02$. Tests are applied
to see that this is reasonable. We restrict ourselves to the case $N_f=3$.

\subsection{$\chi$QCD}

The $\chi$QCD staggered lattice action incorporates an irrelevant chiral
4-fermion interaction of the Gross-Neveu form, which allows us to run at zero
quark mass. In terms of pseudo-fermion fields $\psi$ and their conjugate momenta
$p_\psi$ the fermion action, modified for implementation of the RHMC algorithm
is
\begin{equation}
S_{pf}=p_\psi^\dag (A^\dag A)^{-N_f/8} p_\psi
\end{equation}
where
\begin{equation}
A = \not\!\! D + m + \frac{1}{16} \sum_i (\sigma_i+i\epsilon\pi_i)
\end{equation}
with $i$ running over the 16 sites on the dual lattice neighbouring the site
on the normal lattice, $\epsilon=(-1)^{x+y+z+t}$ and $\not\!\! D$ is the usual
gauge-covariant ``d-slash'' for the staggered quarks. $\sigma$ and $\pi$ are
the auxiliary fields which implement the 4-fermion interactions. The action
for these auxiliary fields is
\begin{equation}
S_{\sigma\pi} = \sum_{\tilde{s}}\frac{1}{8}N_f\gamma(\sigma^2+\pi^2),
\end{equation}
where the sum is over the sites of the dual lattice. For more details of the
rest of the action we refer the reader to our earlier work
\cite{Kogut:1998rg}. We did, however make one minor change, multiplying the
$(\sigma,\pi)$ `kinetic energy' by $\gamma$, which acts as a large mass and
slows down the dynamics of these chiral fields.

Although we do not know the upper bound of the spectrum of the Dirac operator,
because of the interaction with the chiral field, we know from our HMD(R)
simulations that $\langle\frac{1}{8}N_f\gamma(\sigma^2+\pi^2)\rangle$ is 
just above 1. For $N_f=2$ and $\gamma=10$ the coefficient 
$\frac{1}{8}N_f\gamma=2.5$. This plus the fact that the $\sigma$ and $\pi$
interacting with the fermion fields at a given point are averages over 16
points on the dual lattice, suggest that 1 is a conservative upper bound for
the contribution of the chiral fields to the eigenvalues of the Dirac operator.
We thus assume an upper bound of $25$ for the spectrum of the quadratic
Dirac operator. If this were violated, it would only affect modes close to
the ultraviolet cutoff on the theory, which are of limited interest. We shall
later present evidence that the magnitude of the chiral field which interacts
with the fermion field at a point {\it is} bounded by $1$.

More important is an accurate knowledge of the lower bound on the spectrum of
the quadratic Dirac operator. Here we simply assume that the action of the
chiral fields on the spectrum is similar to giving the quarks a small mass.
Our speculative lower bound is that for a regular staggered action with a quark
mass $0.001$, i.e. $1 \times 10^{-6}$, and we perform tests of this assumption.

\section{Tests of the RHMC for lattice QCD at finite $\mu_I$.}

The RHMC algorithm for QCD at a finite chemical potential $\mu_I$ for isospin
is implemented following Clark and Kennedy \cite{clark3}. 
The rational approximations are
obtained using the Michael Clark's Remez code \cite{clark2}. 
We have been using this method
for simulating $N_f=3$ flavour QCD. Here we use a $(20,20)$ rational 
approximation to ${\cal M}^{(3/16)}$ needed at the beginning of each trajectory,
and a $(20,20)$ rational approximation to ${\cal M}^{(-3/16)}$ needed at the
end of the trajectory to calculate the final `energy' (value of the classical
molecular-dynamics Hamiltonian describing the system), over the interval
$[1 \times 10^{-4},25]$ discussed in section~2. These approximations have a
maximum relative error of $6.1 \times 10^{-12}$. The evolution of the fields
over the trajectories were performed using a $(10,10)$ rational approximation
to ${\cal M}^{(-3/8)}$ over the interval $[1 \times 10^{-4},25]$, which has 
maximum relative error of $4.4 \times 10^{-6}$. Since we have been using
relatively small lattices $8^3 \times 4$ and $12^3 \times 4$, we have found
32-bit floating-point precision (accumulations are performed in 64-bit 
precision) to be adequate. 

For the RHMC algorithm to be exact, i.e. for it to produce an ensemble of
configurations which have the correct Boltzmann distribution for the given
action with no $dt$ dependence, the updating algorithm must be reversible. The
updating method specified by Kennedy et al. is reversible for infinite
precision arithmetic, even when the multimass conjugate gradient solver used
to invert the Dirac operator is stopped prior to convergence. (The inversions
at the beginning and end of each trajectory still need to be exact for this to
hold). With the finite precision used by computers, this reversibility must be
tested. As mentioned above, we use single precision (32-bit) floating point
arithmetic for simulations of lattice QCD at finite $\mu_I$. To test this we
performed a 1000 trajectory run on a $12^3 \times 4$ lattice at
$\beta=5.1285$, $m=0.03$ and $\mu_I=0.3$, in which every second trajectory was
run forward and then reversed, returning to its initial state. $dt=0.05$ for
this experiment, and the trajectory length $\Delta t=1$. $\mu_I=0.3$ is
chosen, because it is the largest $\mu_I$ used in our production runs.
$m=0.03$, just above the critical mass, is the smallest mass for which we run
at $\mu_I=0.3$, and $\beta=5.1285$, close to the transition $\beta$ for these
values of $m$ and $\mu_I$, is one we use for production runs. We expect
therefore that the Dirac operator will be as close to singular as we will
encounter in any of our simulations, and that if the algorithm is satisfactory
here, it will work for the whole range of parameters we use.

\begin{figure}[htb]
\epsfxsize=4in
\centerline{\epsffile{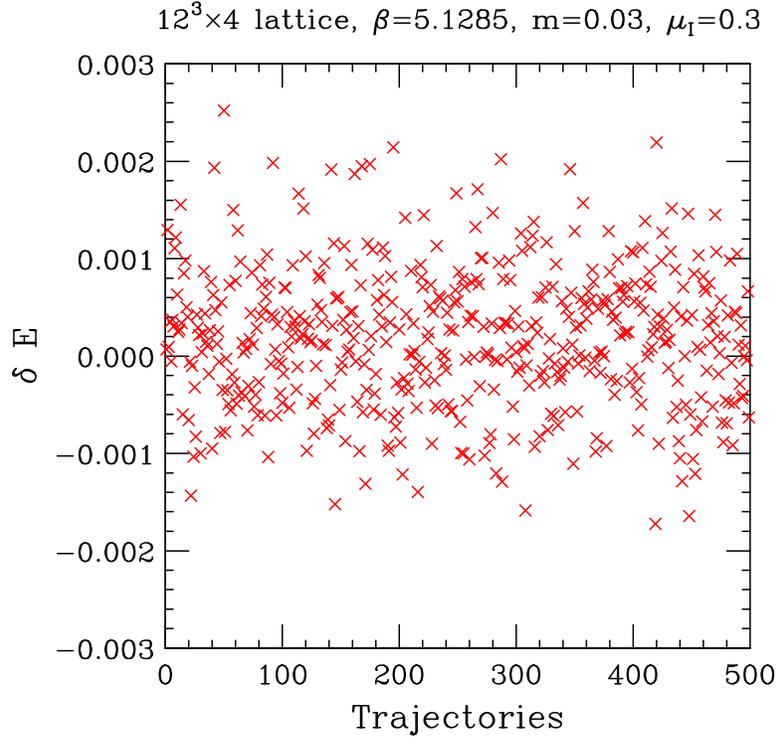}}
\vspace{0.2in}
\centerline{\epsffile{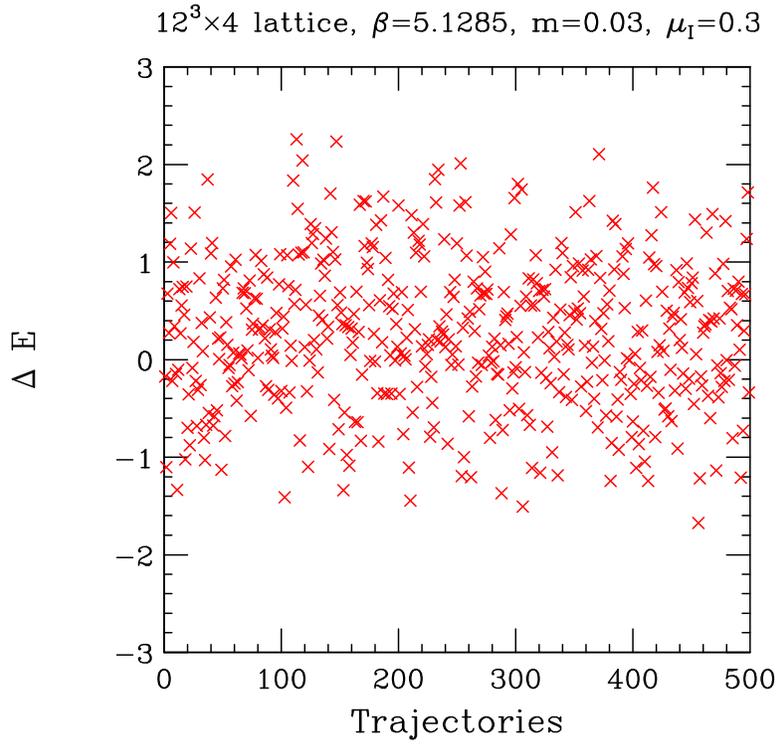}}
\caption{a) The energy change for `closed' trajectories 
($t \rightarrow t+1 \rightarrow t$). b) The energy change for `open' 
($t \rightarrow t+1$) trajectories.}
\label{fig:dEiso}
\end{figure}

The change in `energy' $\delta E$, for each of the 500 reversed trajectories
is plotted in figure~\ref{fig:dEiso}a. The change in energy for each of the
non-reversed trajectories is shown in figure~\ref{fig:dEiso}b.
We choose $E$ as a measure of reversibility because it is the quantity that is
used in the global Metropolis accept reject step. Hence departures of $\delta E$
from zero, its value for true reversibility, indicate the size of errors
introduced from departures from exact reversibility coming from finite precision
errors potentially magnified by instabilities. As we see, these are indeed small
in magnitude -- $< 3 \times 10^{-3}$, and $3$ orders of magnitude smaller than
the change $\Delta E$ in energy over a typical trajectory used for normal
updating. In addition, the average $\delta E$ over the 500 trajectories
considered is only $\approx 1.8 \times 10^{-4}$, indicating that there is no 
sign of a systematic bias, so that the small errors introduced by this lack of 
reversibility are likely to cancel. 

Most of the simulations of lattice QCD at finite isospin density, which we 
have run using this code use $dt=0.05$, and all use trajectory length 
$\Delta t=1$. With this choice we find an acceptance rate of $\sim 70\%$ for
the generated trajectories. Most of our runs used the same $\beta$ for the
updating as for the global Metropolis step, since we did not find large gains
from choosing different $\beta$s.

We check our speculative lower bound on the spectrum of the quadratic 
Dirac operator of $1 \times 10^{-4}$, at our largest $\mu_I$ ($\mu_I=0.3$)
at both masses ($0.03$ and $0.035$) where we performed production simulations.
This we did by running using rational approximations, valid over intervals
with the same upper bound, but different lower bounds, and compared observables.
At $m=0.035$, $\mu_I=0.3$ and $\beta=5.1370$, we performed a 300,000 trajectory
run with rational approximations valid over the interval 
$[1 \times 10^{-4},25]$ used for production running and a 300,000 trajectory
run with the same rational approximation for updating, but a $(20,20)$ rational
approximation valid over the range $[1 \times 10^{-5},25]$ for initiating
each trajectory and for calculating the energy at the end of each trajectory.
This approximation has maximum relative error of $2.0 \times 10^{-10}$ . A 
comparison of the observables from these 2 simulations is given in
table~\ref{tab:compare}a. At $m=0.03$, $\mu_I=0.3$, $\beta=5.1290$ we ran
one 300,000 trajectory run with rational approximations valid over our default
interval ($[1 \times 10^{-4},25]$) and one 300,000 trajectory run with rational
approximations valid over the interval $[1 \times 10^{-6},25]$ for 
initialization, measurement {\it and} updating each trajectory. For this later
interval we used $(25,25)$ rational approximations with maximum relative error
$2.0 \times 10^{-11}$ at the ends of each trajectory and $(15,15)$ rational
approximations with a maximum relative error of $7.2 \times 10^{-7}$ for the
updating. Observables from these 2 runs are compared in 
table~\ref{tab:compare}b.
\begin{table}[htb]
\begin{tabular}{lll}
\multicolumn{3}{c}{$m=0.035$, $\mu_I=0.3$, $\beta=5.1370$}                \\
\hline
Spectrum range  &   $[1 \times 10^{-4},25]$   &   $[1 \times 10^{-5},25]$ \\
\hline
$B_4$                           &   1.860(36)          &   1.879(41)      \\
$\beta_c$                       &   5.13744(24)        &   5.13787(22)    \\
plaquette                       &   0.51086(66)        &   0.51134(60)    \\
Wilson Line                     &   0.3300(74)         &   0.3245(68)     \\
$\langle\bar{\psi}\psi\rangle$  &   0.6274(94)         &   0.6347(86)     \\
$j_0^3$                         &   0.0450(10)         &   0.0442(10)     \\
\hline
\end{tabular}

\vspace{0.2in}

\begin{tabular}{lll}
\multicolumn{3}{c}{$m=0.03$, $\mu_I=0.3$, $\beta=5.1290$}                 \\
\hline
Spectrum range  &   $[1 \times 10^{-4},25]$   &   $[1 \times 10^{-6},25]$ \\
\hline
$B_4$                           &   1.704(30)         &   1.722(39)       \\
$\beta_c$                       &   5.12826(18)       &   5.12820(21)     \\
plaquette                       &   0.50776(60)       &   0.50753(68)     \\   
Wilson Line                     &   0.3773(68)        &   0.3792(76)      \\
$\langle\bar{\psi}\psi\rangle$  &   0.5442(80)        &   0.5411(103)     \\
$j_0^3$                         &   0.0529(10)        &   0.0532(11)      \\
\hline  
\end{tabular}

\caption{Comparison of observables measured during RHMC simulations with 
different speculative lower bounds for the spectrum of the quadratic Dirac
operator: a) For $m=0.035$, $\mu_I=0.3$, $\beta=5.1370$. 
b) For $m=0.03$, $\mu_I=0.3$, $\beta=5.1290$.}
\label{tab:compare}

\end{table}

The good agreement between the `data' produced with the different values
for the speculative lower bound indicates that our initial choice of
$1 \times 10^{-4}$ was adequate. One notes that, despite the high statistics --
300,000 trajectories for each run, the error bars are sizable. This is 
because the chosen $\beta$ values are very close to $\beta_c$, the transition
$\beta$, in each case which serves to maximize the fluctuations.

Finally we compare the values of two fluctuation quantities, the Binder 
cumulant $B_4(\bar{\psi}\psi)$ and the chiral susceptibility
$\chi_{\bar{\psi}\psi}$ at $\beta_c$ with results obtained from our HMD(R)
simulations over a range of $dt$ values for $m=0.035$ and an intermediate
value of $\mu_I$, $\mu_I=0.2$, in figure~\ref{fig:B4,chi}. The leading
errors in HMD(R) are expected to be ${\cal O}(dt^2)$. The fits linear in
$dt^2$ in these figures appear to support this expectation, but indicate that
we would need to include an ${\cal O}(dt^4)$ term to fit over the whole 
range of measurements. We note that the finite $dt$ errors in both these
graphs are quite large, so that agreement between HMD(R) and RHMC is 
non-trivial.  

\begin{figure}[htb]
\epsfxsize=4in
\centerline{\epsffile{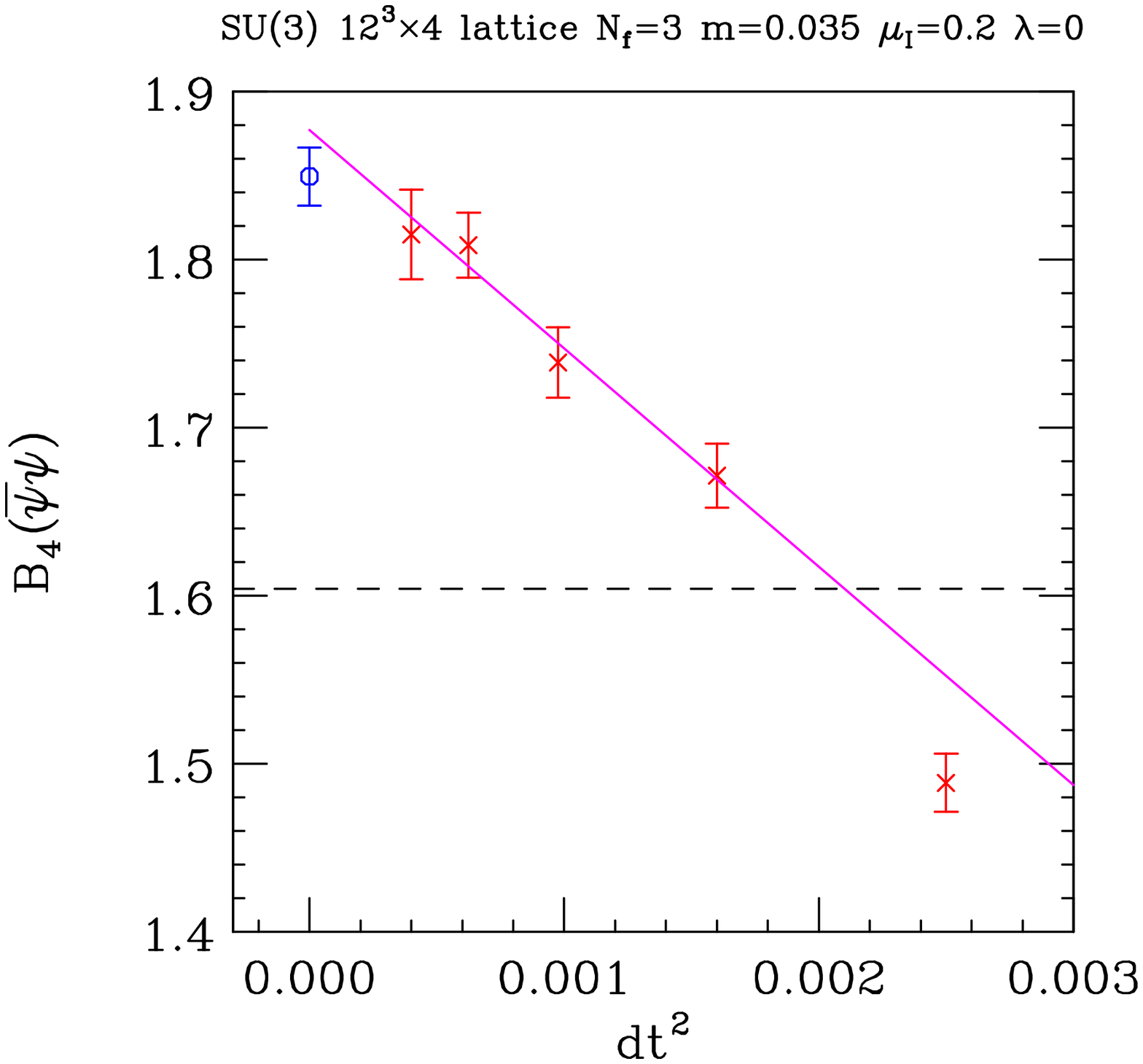}}
\vspace{0.2in}
\centerline{\epsffile{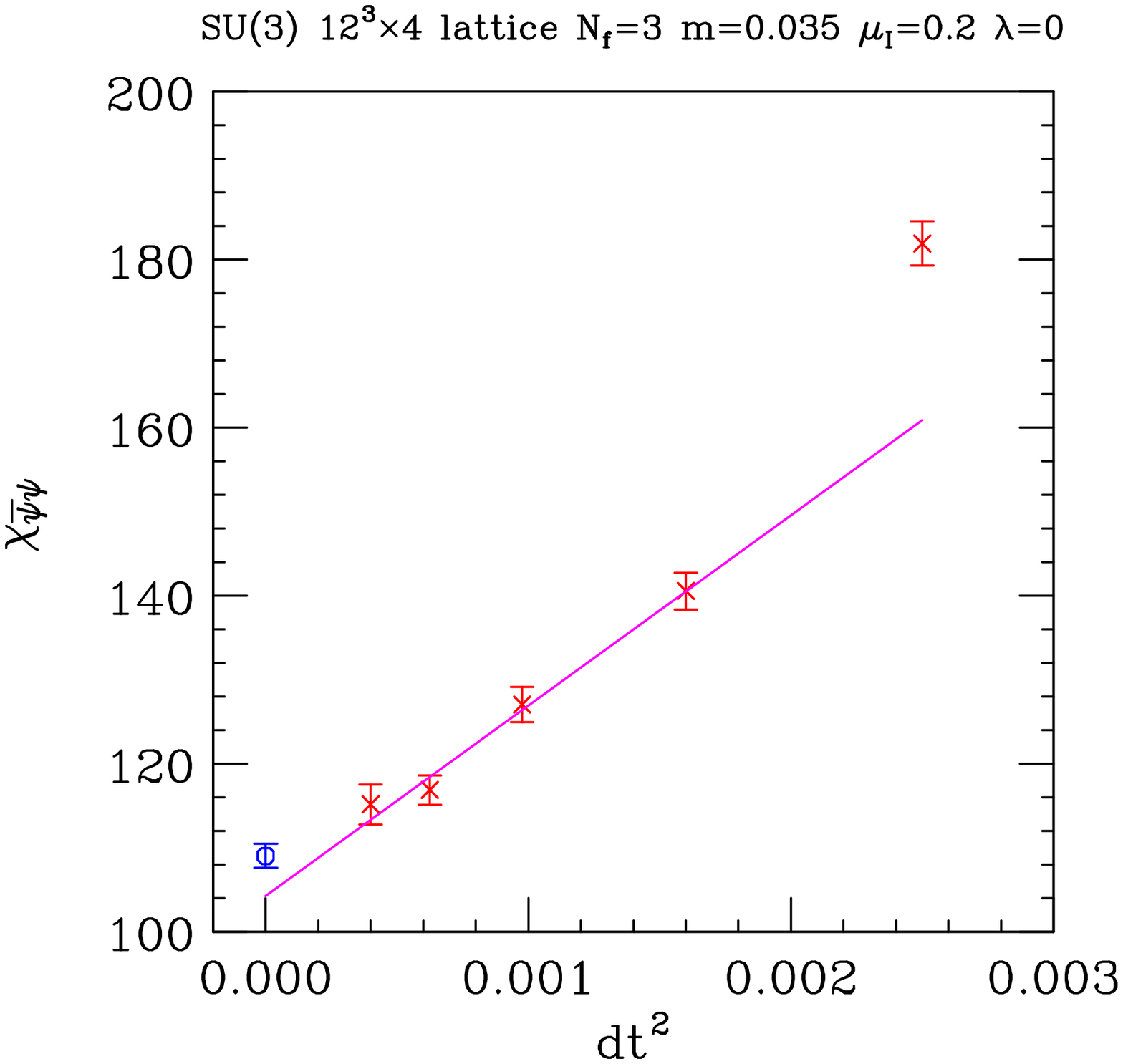}}
\caption{a) The Binder cumulant $B_4(\bar{\psi}\psi)$ as a function of
$dt^2$ for the HMD(R) algorithm (crosses) and the RHMC algorithm (circle at
$dt=0$). b) The chiral susceptibility $\chi_{\bar{\psi}\psi}$ as a function of
$dt^2$ for the HMD(R) algorithm (crosses) and the RHMC algorithm (circle at
$dt=0$).}
\label{fig:B4,chi}
\end{figure}

\section{Tests of RHMC for $\chi$QCD}

Since we have been simulating the thermodynamics of $\chi$QCD with 2 massless 
quark flavours on $16^3 \times 8$ and $24^3 \times 8$ lattices, we use this
larger lattice to run our tests. On this size lattice, extensive comparisons
with the HMD(R) algorithm over a range of $dt$ values is impractical, as is
performing high statistics runs with various speculative lower bounds.

Our first tests used a single-precision (32-bit floating-point) version of the
code, then a modified version where the multimass conjugate gradient
inversions which initiate and end each trajectory were replaced with double-%
precision (64-bit floating-point) versions, while leaving the inversion
routines which perform the updates within each trajectory in single precision.
We then performed reversibility tests of the same nature as those described in
the previous section. For the updating, our convergence criterion was applied
to the normalized conjugate gradient residual for the lowest lying pole in the
partial fraction expansion of the rational approximation, while at the
beginning and end of each trajectory it was applied to the unnormalized
residual. (If $x_0$ is an approximate solution to the linear equations $Ax=b$,
the unnormalized residual is $r_u=|b-Ax_0|$. The normalized residual is
$r_n=r_u/|b|$.) We denote this upper limit on these residuals by $r$ and note
that, since our sources are extensive, this means that we are using a much
more stringent convergence criterion at the beginning and end of each
trajectory than in the updating, as required.

\begin{figure}[htb]
\epsfxsize=4in
\centerline{\epsffile{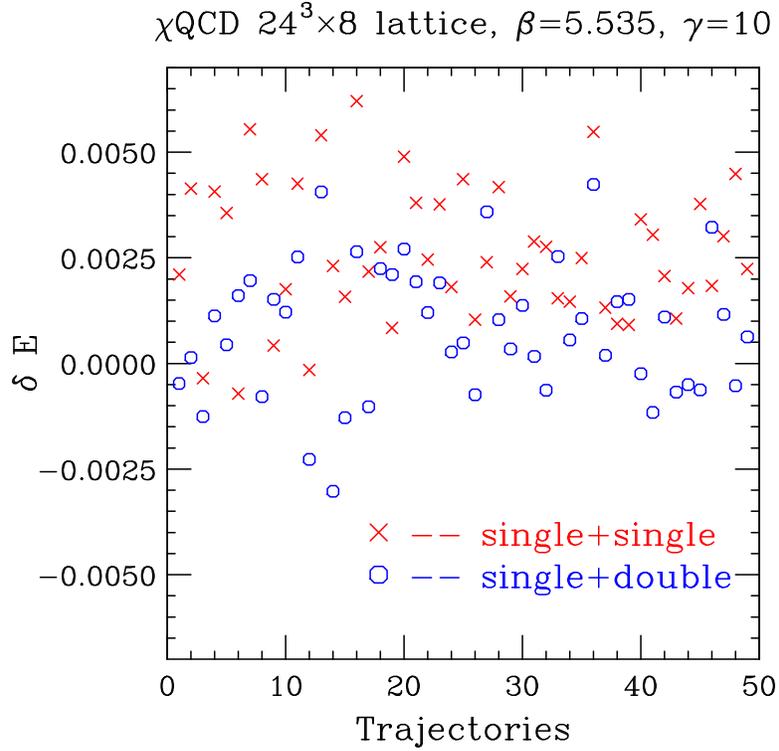}}
\vspace{0.2in} 
\centerline{\epsffile{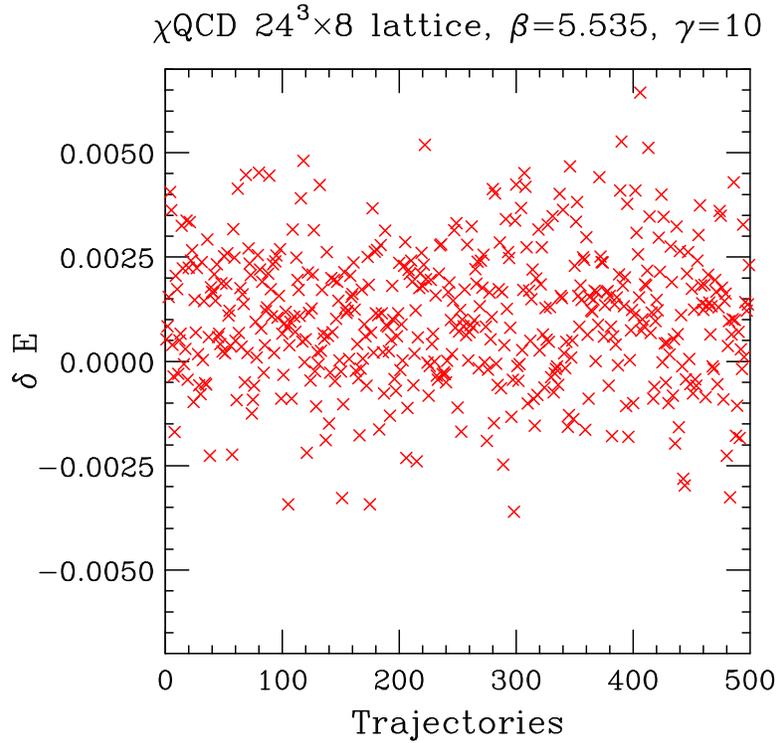}}
\caption{a) Change in energy $\delta E$ for 50 closed trajectories for single
precision code (single-single) and single precision updating with double
precision initialization and measurement for each trajectory (single-double).
b) 500 closed trajectories for single-double code.}
\label{fig:dEsingle}
\end{figure}

Figure~\ref{fig:dEsingle}a shows the changes in energy over the `closed'
trajectories $t \rightarrow t+1 \rightarrow t$ for the single precision code
and for the same code with the initial and final inversions for each trajectory
performed in double precision. $r=0.001$ for these runs, since we have 
determined that no improvement is obtained by decreasing $r$ below this value.
$\beta=5.535$, used for these simulations, is close to the critical point.
The trajectory length is $1$ time unit and $dt=0.025$ which gives an acceptance
of $\approx 75\%$. However, we turn off the accept/reject step in the RHMC
algorithm for the purpose of these reversibility tests to speed up the passage
of the system through phase space.
The single precision code produces energy changes as large as $\sim 6 \times
10^{-3}$. More troubling is the fact that $\delta E$ is strongly biased to
positive values. In fact the average $\delta E$ is $\approx 2.6 \times
10^{-3}$ which suggests that this code could produce small but significant
systematic errors. The mixed single/double precision code shows more promise
and we have extended our run to include 500 closed trajectories as shown in
figure~\ref{fig:dEsingle}b. The maximum value of $|\delta E|$ is $\approx 6.4
\times 10^{-3}$, which is larger than we would prefer. More problematic is
that the average $\delta E$ over the range is $\approx 1.1 \times 10^{-3}$,
which suggests a systematic bias as is apparent in figure~\ref{fig:dEsingle}b.
Unfortunately, we are limited by precision from improving this situation. We
thus feel that to be safe we should increase the precision, especially if we
wish to allow ourselves the option of increasing the lattice size in the future.

\begin{figure}[htb]
\epsfxsize=6in
\centerline{\epsffile{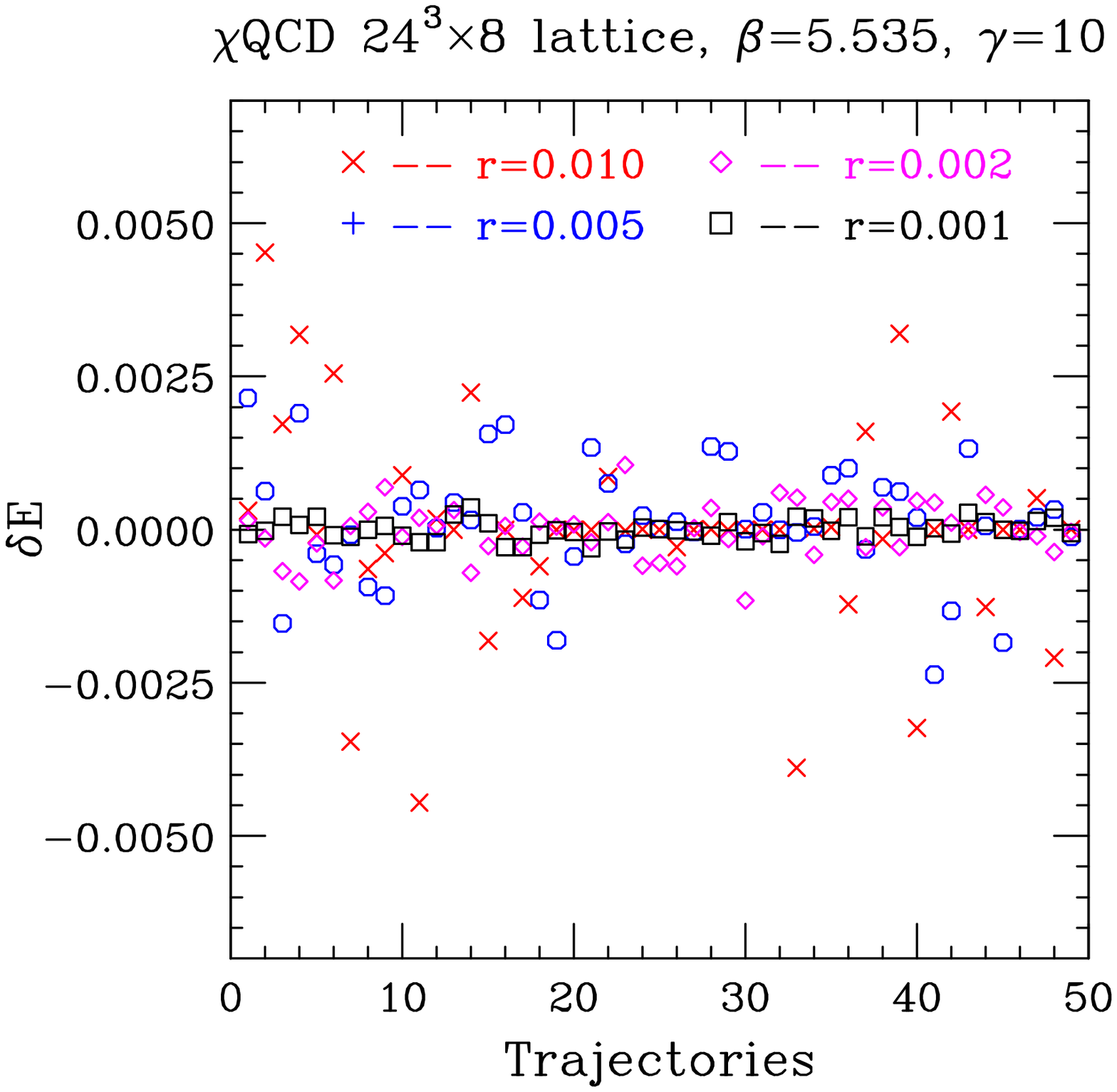}}
\caption{The change in energy $\delta E$ for closed trajectories for a range
of choices of $r$ for 64-bit precision code.}
\label{fig:dEdouble}
\end{figure}

\begin{figure}[htb]
\epsfxsize=4in
\centerline{\epsffile{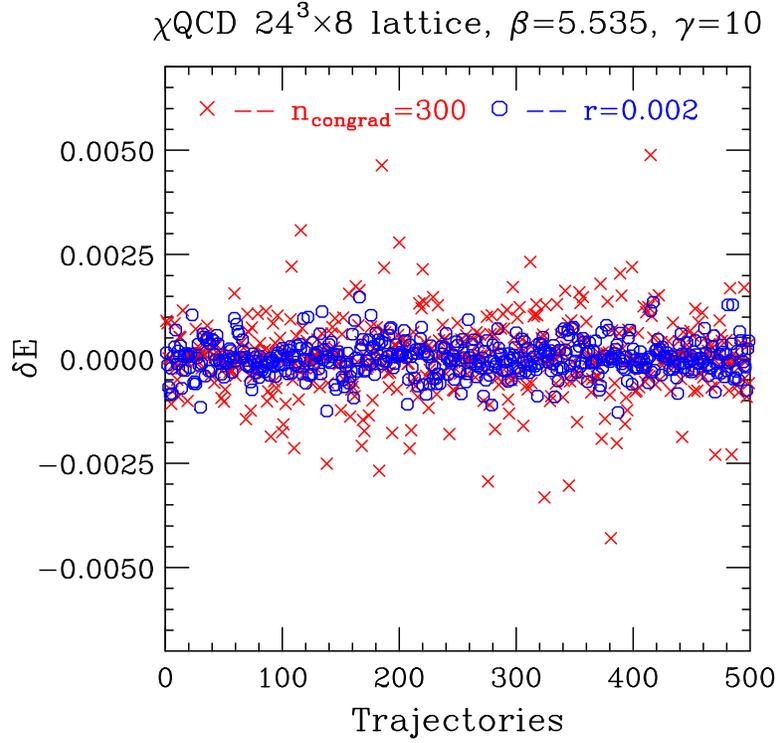}}
\vspace{0.2in}
\centerline{\epsffile{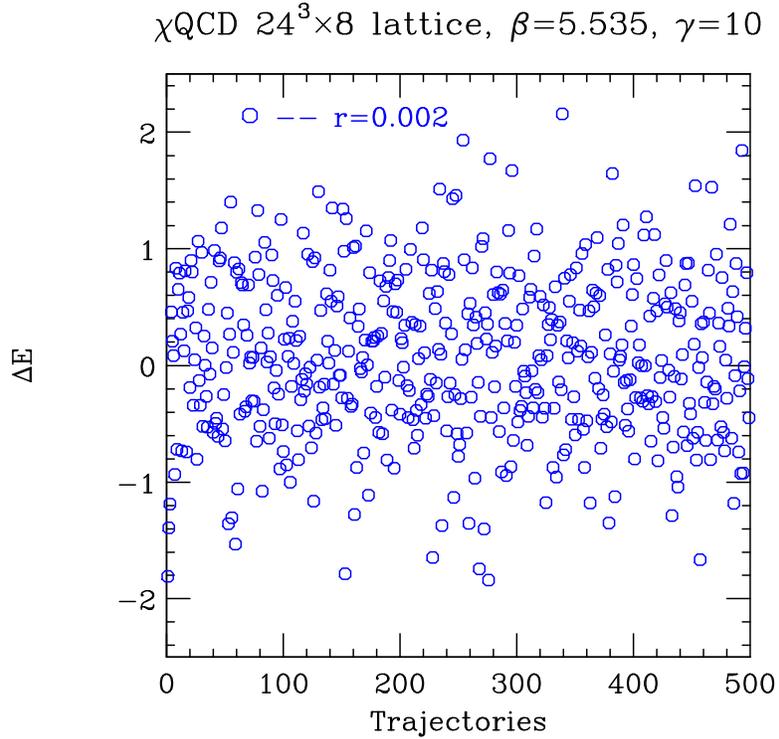}}
\caption{a) The change in energy $\delta E$ for closed trajectories for a fixed
number of conjugate gradient iterations and for $r=0.002$. b) The change in
energy $\Delta E$ for open trajectories for $r=0.002$.}
\label{fig:dEdouble2}
\end{figure}

We have tested reversibility of our double precision code as a function of
the residual $r$. The results for 50 closed trajectories are shown in 
figure~\ref{fig:dEdouble}.
We see that this measure of reversibility of the trajectories improves 
considerably as $r$ is decreased from $0.01$ to $0.001$. Comparing this
graph with that for the single precision tests, we see that precision is indeed
the limiting factor. Further tests indicate that it is the convergence of the
approximate inversions during the updating (rather than that of the more
precise inversions at the start and end of each trajectory) that is the
limiting factor. Careful monitoring of the trajectories for the $r=0.01$
run, indicates that this is largely due to the fact that the number of 
iterations of the conjugate gradient inverter sometimes differed by 1 or 2
iterations between the corresponding steps of the forward and reversed 
trajectories, and this of course has the largest effect when one is far from
convergence. This can be corrected by running for a fixed number of conjugate
gradient iterations rather than at fixed $r$. Since at $r=0.01$ it requires
around 250 iterations on average for the inversion at each update, we performed
a run of 1000 trajectories with the number of inverter iterations at each
update set to 300 (those at the ends of the trajectories were still set by a
conservative choice of $r$). The results of running for 1000 trajectories (500
closed) with the number of conjugate gradient iterations fixed at 300 are
shown in figure~\ref{fig:dEdouble2}a along with an identical number of
trajectories with these inversions set by $r=0.002$. The maximum value of
$|\delta E|$ for the case where the number of conjugate gradient iterations is
fixed at 300 is $\sim 5 \times 10^{-3}$, no better than the single precision
results. For $r=0.002$ the maximum value of $|\delta E|$ is $\sim 1.5 \times
10^{-3}$ which we consider acceptable. The reason that fixing the number of
conjugate gradient iterations at a relatively small value did not give better
results is presumably because this gives relatively poor convergence for some
configurations. Studies of the HMC algorithm have indicated that if the
convergence is too poor, the updating procedure develops instabilities which
amplify the effects of finite precision errors \cite{Joo:2000dh}. We believe
that such instabilities are most likely the reason why using a fixed number of
conjugate gradient iterations during updating did not lead to smaller
violations of reversibility. Figure~\ref{fig:dEdouble2}b shows the change of
energies for the open trajectories from the same $r=0.002$ run.

\begin{figure}[htb]
\epsfxsize=6in                                                                
\centerline{\epsffile{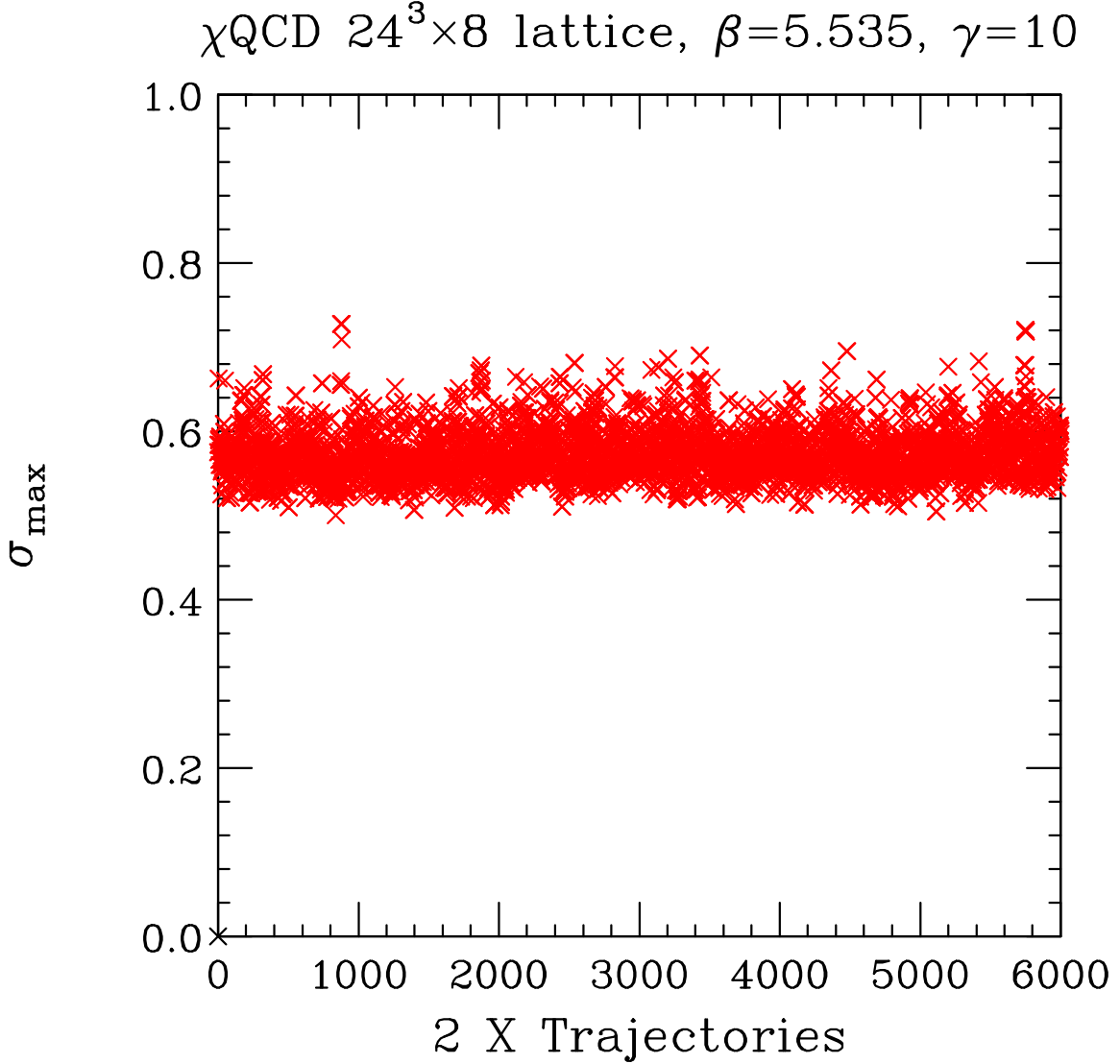}}
\caption{Maximum values of $|\sigma+i\pi|$ at the start and end of each
trajectory of a production run.}
\label{fig:sigma}
\end{figure}

We now need to justify our choices of spectral bounds on the quadratic Dirac
operator. The chosen upper bound of $25$ is obtained assuming that the
magnitude of the average of the 16 chiral auxiliary fields $\sigma+i\pi$ which
couple to the fermion bilinear $\bar{\psi}\psi$ at each site is bounded above
by $1$. Figure~\ref{fig:sigma} shows the maximum over the lattice sites of
the magnitude of $\sigma+i\pi$ at the beginning and end of 3000 trajectories
of our production running at $m=0$, $\beta=5.535$, $\gamma=10$, $dt=0.03125$,
trajectory length $1$ and $r=0.002$, where the acceptance is around 60--70\%.
As can be seen, $1$ {\it is} a conservative upper bound for $|\sigma+i\pi|$.

Since we are using a relatively large lattice ($24^3 \times 8$) it is not
practical to obtain adequate statistics close to the phase transition with
more than one choice of speculative lower bounds as was done with our
simulations of lattice QCD at finite isospin chemical potential, in order
to check for consistent results. Instead, we performed a simulation of 1000
trajectories with a speculative lower bound of $1 \times 10^{-8}$. At the
beginning of each trajectory where we calculate the momenta $p_\psi$ conjugate
to the pseudo-fermion fields, as
\begin{equation}
p_\psi = {\cal M}^{1/8} \xi,
\end{equation}
we compared the values of $p_\psi$ obtained with rational approximations
which assume lower bounds of $1 \times 10^{-6}$, $1 \times 10^{-8}$ and
$1 \times 10^{-10}$. These used $(25,25)$, $(30,30)$, and $(35,35)$ rational
approximations respectively, with maximum relative errors of 
$1.4 \times 10^{-11}$, $3.0 \times 10^{-11}$ and $4.9 \times 10^{-11}$
respectively. Similar comparisons were made of the $\xi$s at the end of
each trajectory reconstituted as
\begin{equation}
\xi = {\cal M}^{-1/8} p_\psi.
\end{equation}
No comparisons with results from using different speculative lower bounds were
made during the updates, since using different rational approximations
corresponds to making different choices of the updating Hamiltonian, and such
choices are unimportant unless they significantly affect the acceptance.

\begin{figure}[htb]
\epsfxsize=4in
\centerline{\epsffile{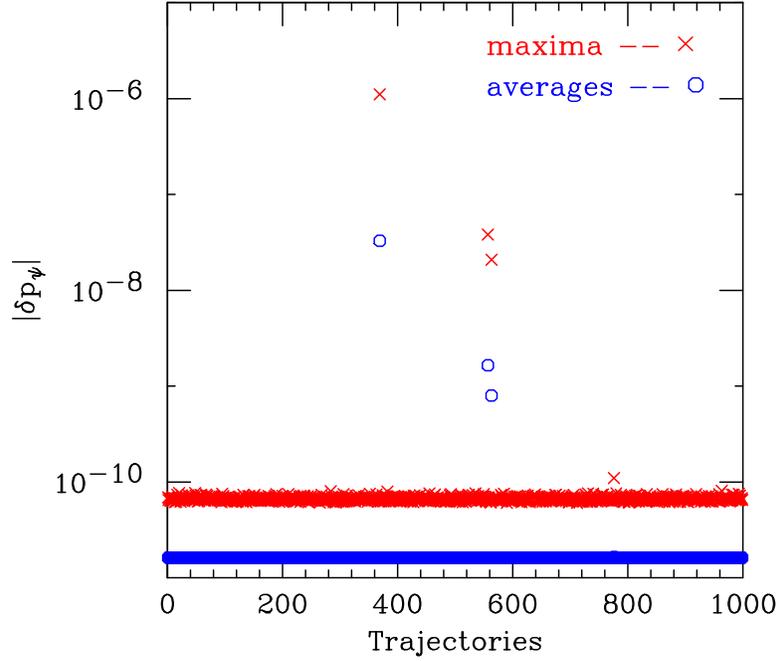}}
\vspace{0.2in}
\centerline{\epsffile{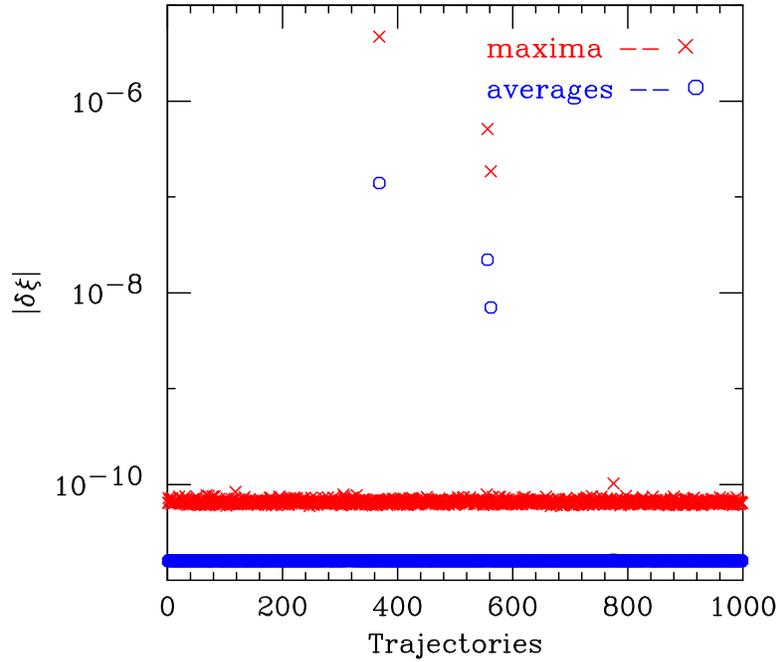}}
\caption{a) Magnitude of the difference between $p_\psi$ values using 
speculative lower bounds of $10^{-6}$ and $10^{-8}$. b) Magnitude of the 
difference between $\xi$ values using speculative lower bounds of $10^{-6}$ 
and $10^{-8}$.}
\label{fig:68}
\end{figure}

\begin{figure}[htb]
\epsfxsize=4in 
\centerline{\epsffile{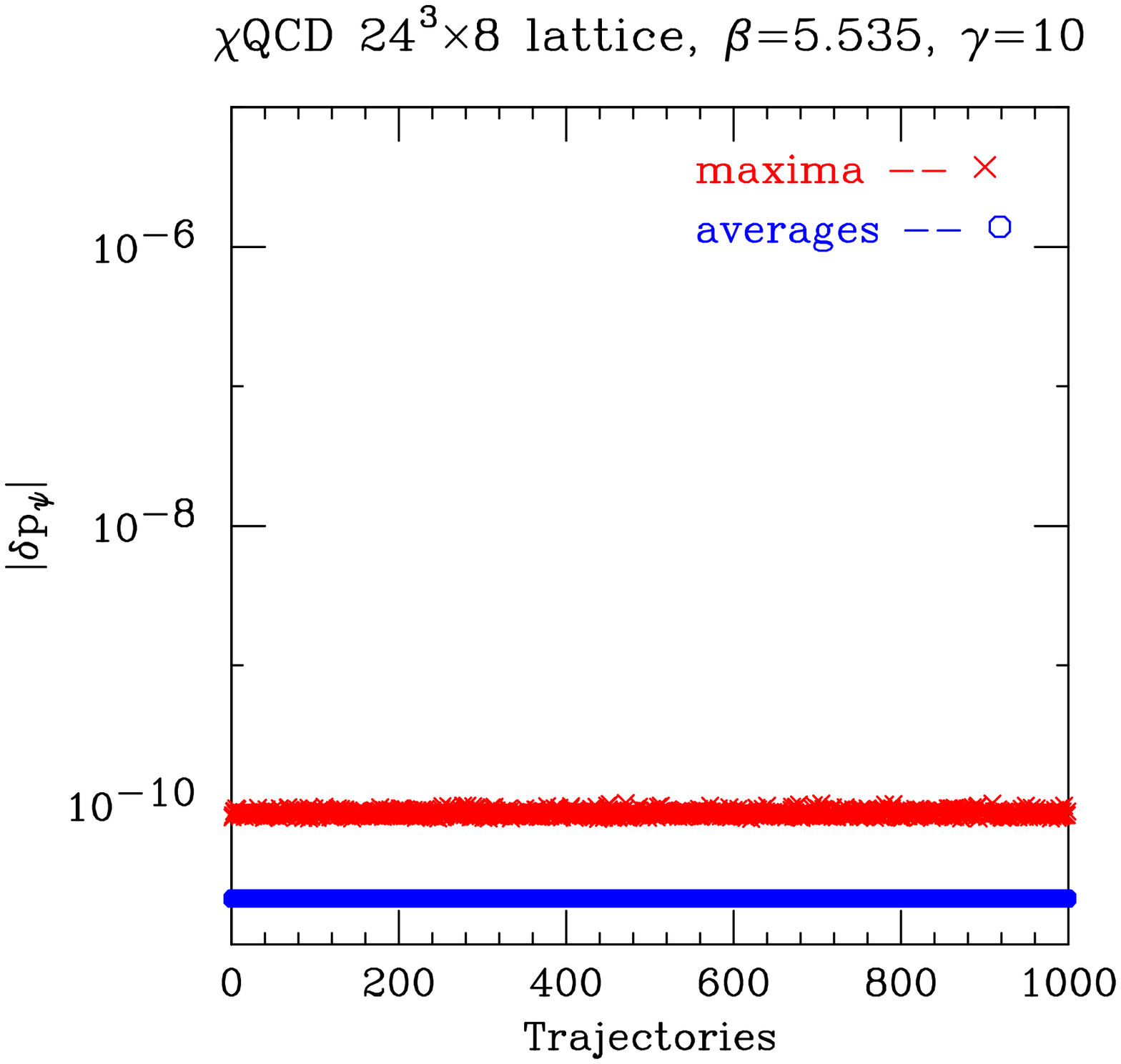}}
\vspace{0.2in}                    
\centerline{\epsffile{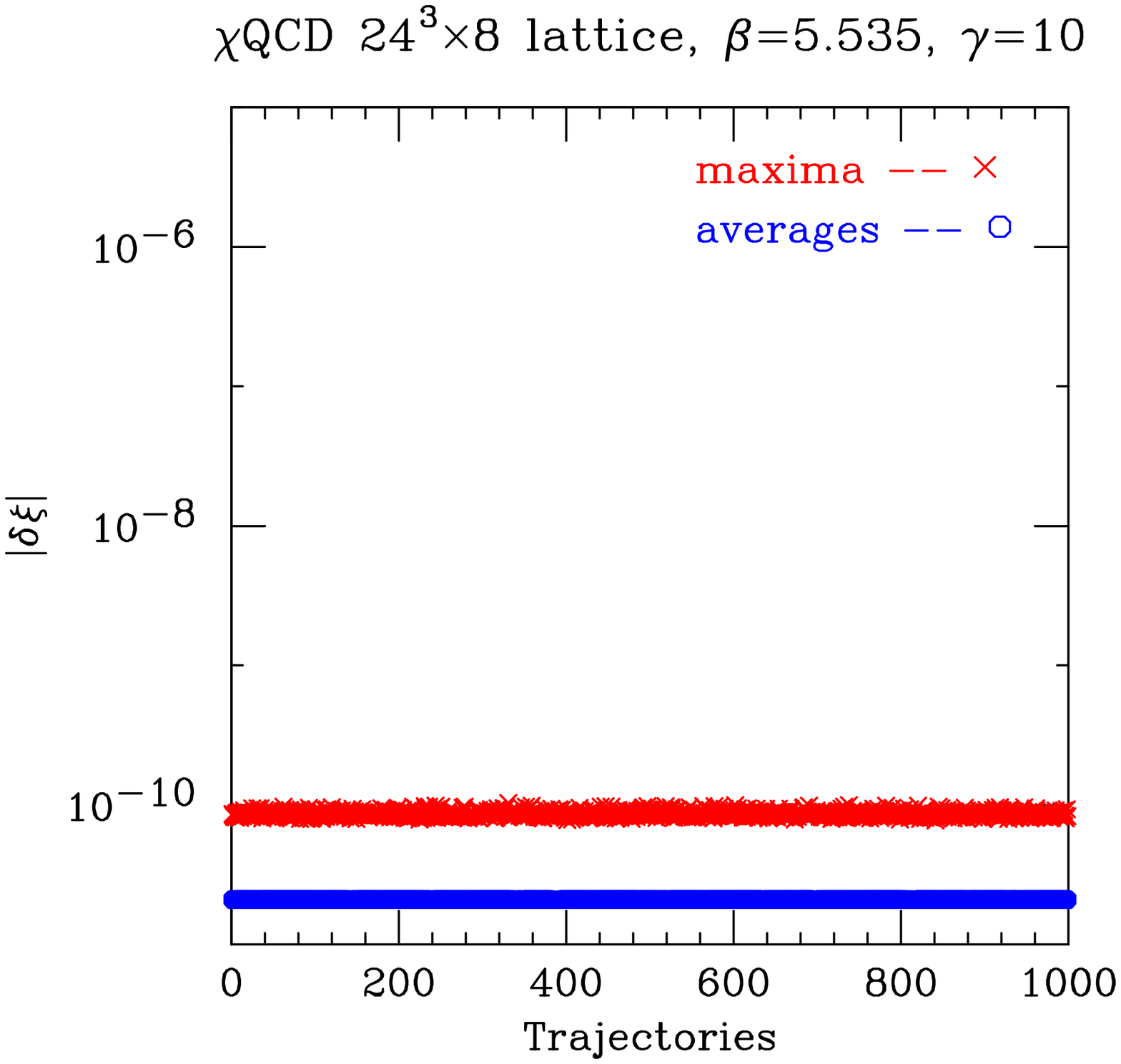}}
\caption{a) Magnitude of the difference between $p_\psi$ values using
speculative lower bounds of $10^{-10}$ and $10^{-8}$. b) Magnitude of the
difference between $\xi$ values using speculative lower bounds of $10^{-10}$
and $10^{-8}$.}                        
\label{fig:810}
\end{figure}

In figure~\ref{fig:68}a, we show the magnitude of the difference between
$p_\psi$ values calculated using a lower bound of $1 \times 10^{-6}$ and
those using a lower bound of $1 \times 10^{-8}$. We plot the maximum over sites
and colours of this quantity for each trajectory as well as the average over
sites and colours. Figure~\ref{fig:68}b shows such differences for $\xi$.
For 997 of these trajectories, the differences in $p_\psi$, and $\xi$ values
are consistent with zero within the accuracies of the rational approximations 
we use. For the other 3 trajectories, the differences are much larger. In 
addition, we note that the configurations which give relatively large errors 
for $p_\psi$, and for $\xi$, are identical. Figure~\ref{fig:810} shows 
differences between the same quantities for lower bounds of $10^{-8}$ and 
$10^{-10}$ respectively for the {\it same} trajectories as in 
figure~\ref{fig:68}.
In this case all differences are consistent with zero within the accuracy
of the rational approximations we use.

We interpret these results as indicating that for 997 of the configurations
generated, all eigenvalues of the quadratic Dirac operator are 
$\gtrsim 10^{-6}$. For the remaining 3 configurations, the minimum eigenvalues
lie between $10^{-8}$ and $10^{-6}$. It is reassuring to note that even for
these configurations the differences are small, since a typical $p_\psi$ or
$\chi$ has a magnitude ${\cal O}(1)$. Hence it is probably reasonable to use a
speculative lower bound of $10^{-6}$ and, unless this run was anomalous, 
choosing a speculative lower bound of $10^{-8}$ is completely safe.

Finally we present some results from small lattice ($8^3 \times 4$) simulations
using the RHMC algorithm for $\chi$QCD and compare them with earlier published
results using the HMD(R) algorithm \cite{Kogut:1998rg}. 
For such small lattices precision is not
an issue and these simulations were performed using 32-bit floating-point
arithmetic. Table~\ref{tab:rhmc}a gives the results for $\gamma=10$, $m=0$ for
our RHMC simulations over a range of $\beta$ values which bracket the phase
transition. We ran 100,000 trajectories for each $\beta$. Table~\ref{tab:rhmc}b
gives the published results for the same parameters for the HMD(R) algorithm.
\begin{table}[htb]
\begin{tabular}{ddddd}
\hline
\multicolumn{1}{c}{$\beta$} & 
\multicolumn{1}{r}{Wilson Line} & 
\multicolumn{1}{@{\hspace{0.5in}}c}{$\langle\bar{\psi}\psi\rangle$} & 
\multicolumn{1}{@{\hspace{0.5in}}c}{$\langle\sigma\rangle$} &
\multicolumn{1}{@{\hspace{0.5in}}c}{Plaquette}        \\
\hline
5.0     & 0.0253(3)  &  1.2192(4)  & 0.1227(2) & 0.58256(5)               \\   
5.1     & 0.0300(4)  &  1.1556(6)  & 0.1167(3) & 0.56450(7)               \\
5.2     & 0.0397(5)  &  1.0620(8)  & 0.1071(3) & 0.54258(8)               \\
5.3     & 0.0832(17) &  0.9008(20) & 0.0916(4) & 0.51422(16)              \\ 
5.4     & 0.6006(12) &  0.1328(19) & 0.0205(3) & 0.46210(6)               \\
5.5     & 0.6838(8)  &  0.0946(13) & 0.0178(3) & 0.44672(5)               \\
\hline
\end{tabular}
\caption{a) Observables measured in RHMC simulations on an $8^3 \times 4$
with $\gamma=10$, $m=0$.}
\vspace{0.2in}
\begin{tabular}{ddddd}
\hline                                                                         
\multicolumn{1}{c}{$\beta$} &
\multicolumn{1}{@{\hspace{0.5in}}r}{Wilson Line} &
\multicolumn{1}{@{\hspace{0.5in}}c}{$\langle\bar{\psi}\psi\rangle$} &
\multicolumn{1}{@{\hspace{0.5in}}c}{$\langle\sigma\rangle$} &
\multicolumn{1}{@{\hspace{0.5in}}c}{Plaquette}        \\
\hline                                                                         
5.0     & 0.025(2)   & 1.217(3)    & 0.1235(5) & 0.5825(3)                \\ 
5.1     & 0.031(2)   & 1.147(2)    & 0.1170(5) & 0.5636(3)                \\  
5.2     & 0.031(4)   & 1.065(4)    & 0.1080(5) & 0.5430(8)                \\
5.3     & 0.073(7)   & 0.910(5)    & 0.0923(8) & 0.5149(7)                \\
5.4     & 0.595(7)   & 0.131(5)    & 0.0203(8) & 0.4619(4)                \\
5.5     & 0.679(5)   & 0.093(3)    & 0.0173(6) & 0.4469(3)                \\
\hline
\end{tabular}                                                                  
\addtocounter{table}{-1}
\caption{b) Observables measured in HMD(R) simulations on an $8^3 \times 4$ 
with $\gamma=10$, $m=0$.}
\label{tab:rhmc}
\end{table}
Considering the relatively low statistics of the earlier HMD(R) results, the
agreement with the RHMC simulations is excellent and the finite $dt$ errors are
small in this case.

\section{Discussions and Conclusions}

We have implemented the RHMC for 2 lattice actions where the lower bound on the
spectrum of the quadratic Dirac operator is unknown. The first is the action
for staggered lattice QCD with a finite chemical potential $\mu_I$ for isospin,
in the small $\mu_I$ regime. The second is the $\chi$QCD action where the
standard staggered quark action is augmented with a chiral 4-fermion interaction
which allows simulations at zero quark mass. In the first case we simulate
3 fermion flavours, in the second 2. Both situations require taking a fractional
power of the fermion determinant so that standard HMC methods fail, and the
alternative HMD(R) algorithm gives finite $dt$ errors. In each case the RHMC
algorithm is implemented using a speculative lower bound to this spectrum, and
its validity is tested by simulations. This rather naive approach appears 
successful in both cases.

In each case we test the reversibility of our implementation. This would be
exact for infinite precision arithmetic and exact inversions at the beginning
and ending of the trajectories. We test this by reversing our trajectories and 
determining how close we come to the initial state. The measure of this is
the difference between the classical energy of this final configuration and
that of the initial configuration. Since energy differences appear in the
global Metropolis accept/reject step at the end of each trajectory, this
energy difference should be much less than one. Because this energy is an 
extensive quantity, this requirement becomes more stringent as the lattice
size is increased. For QCD at finite $\mu_I$, we are using $12^3 \times 4$
lattices for which we find single-precision (32-bit) floating-point arithmetic 
to be adequate. For $\chi$QCD, where we are simulating on $24^3 \times 8$
lattices double-precision arithmetic appears desirable. Here we have also
studied convergence requirements on the inversions needed to implement
the rational approximations, both at the beginning and end of the trajectory
and in the updating. Those at the beginning and end of each trajectory need
to be performed with high precision. The inversions in the updating can be
performed with less precision. If these updating inversions are too imprecise,
the change in energy will be large and the final configuration will not be
accepted. It appears, however, that the instabilities due to finite precision
are worse when these inversions are too imprecise, and this is more important
in determining the minimum acceptable precision than are acceptance criteria.

For QCD at finite $\mu_I$ we test our speculative lower spectral bounds by
2 methods. First we find that the RHMC results for Binder cumulants and
chiral susceptibilities agree with extrapolations of our HMD(R) results to
$dt=0$, and that the position of the transitions calculated from both types
of simulation appear consistent. Second, at the largest $\mu_I$ we use, we
check that all observables and fluctuation quantities at the transition agree
with those obtained with simulations using speculative lower bounds which are
$10$ or $100$ times smaller than our initial choice. This indicates that a
lower bound of $10^{-4}$ is adequate.

For $\chi$QCD, we find good agreement between measurements made using RHMC
simulations and those made using HMD(R) simulations on $8^3 \times 4$ lattices.
With our production lattice sizes of $24^3 \times 8$, it is impractical to
make such detailed comparisons between the 2 methods. In addition making runs
with sufficient statistics to compare different speculative lower bounds is
also too expensive. Instead, we have compared applying rational approximations 
which assume different lower bounds to a set of 1000 configurations generated
during one of our runs. We find agreement to within the accuracy of our
rational approximations for speculative lower bounds of $10^{-8}$ and 
$10^{-10}$. An approximation using a speculative lower bound of $10^{-6}$ 
agrees with that using a speculative lower bound of $10^{-8}$ on most of
these configurations. Even in those cases where the two estimates do not agree,
the differences are small enough that the errors introduced in a production
run would probably be negligible. We conclude therefore that a lower bound of
$10^{-8}$ is almost certainly acceptable, and one of $10^{-6}$ is probably
acceptable. We did apply this test to our QCD at finite $\mu_I$ simulations 
comparing speculative lower bounds of $10^{-4}$ and $10^{-6}$. Although there
is no indication of any difference within the precision of our measurements,
the limitation of 32-bit precision prevents us from using this test to make
definitive statements.

We conclude that the use of speculative lower bounds for the spectrum of the
quadratic Dirac operator is adequate to allow use of the RHMC exact algorithm
to simulate either of the theories we have considered. For theories where it
is necessary to determine the lower bound dynamically, comparisons between
the results of different speculative lower bounds could be made from a set of
choices starting with that with the largest lower bound, during a run (as we 
have done), and choosing the first approximation which passes the test. This
way the expensive calculations of the rational approximations using the Remez
method, need only be performed once. The order of the rational approximation
required to achieve the desired accuracy with a given lower bound does not
increase very rapidly as that bound is increased. For example, in our 
applications where the relative error is always less than $5 \times 10^{-11}$,
we achieve this for the 2-flavour case using a $(20,20)$ rational
approximation for a lower bound of $10^{-4}$, a $(25,25)$ approximation for
$10^{-6}$, a $(30,30)$ approximation for $10^{-8}$, and a $(35,35)$
approximation for $10^{-10}$. Hence the `library' of rational approximations
one would need to maintain for such an implementation would be rather small.
Such methods could be useful for domain-wall fermions and overlap fermions
which have the problem that their lower spectral bounds are also unknown, and
might need to be determined dynamically \cite{clark3}.

\section*{Acknowledgements}
We thank Philippe de Forcrand and Owe Philipsen for helpful discussions.
The computations we have reported were performed on Bassi and Jacquard at
NERSC, Tungsten and Cobalt at NCSA, DataStar at NPACI/SDSC and Jazz at the
LCRC at Argonne National Laboratory. Access to the NSF computers was through
an NRAC allocation.

\end{document}